\documentclass[aps,nofootinbib]{revtex4}
\usepackage{graphicx,color,amsmath}
\begin{document}
\title{Some Physics Beyond the Standard Model at Gamma Gamma Colliders}
\footnote{Parallel talk given at Gamma Gamma session in 
the International Conference on 
Linear Colliders, LCWS04, 19-23 April 2004, Paris, France.
}
\author{Kingman Cheung}
\address{
Department of Physics and NCTS, National Tsing Hua
University, Hsinchu, Taiwan, R.O.C. }
\date{\today}

\begin{abstract}
In this talk, I describe a few interesting physics beyond the standard model
that are quite unique for $\gamma\gamma$ collisions, including
large extra dimension model (ADD), warped extra dimension model 
(Randall-Sundrum) model, universal extra dimension model, as well as
techni-pions of technicolor models.
\end{abstract}
\maketitle

\section{Introduction}
There are strong reasons to believe that the standard model (SM) is only 
an effective theory below 1 TeV.  Theoretically, it has some 
shortcomings.  
(i) The SM has too many parameters,
(ii) it is not a real unification of all forces, and 
(iii) the gauge hierarchy problem.
There are other observations indicating that the SM is not
satisfactory.  The most striking evidence is the definite though
small neutrino masses.
Most of us also believe that there should
be dark matter and even dark energy.  It is clear 
that the SM cannot provide these components of the universe.  In
addition, the SM cannot fulfill all the requirements to give a
sufficiently large enough baryon asymmetry of the universe.
All these problems lead us to believe that there should be
new physics beyond the SM, which should come in TeV scale.  

The hierarchy problem has motivated many new physics.
In recent years, a number of models in extra dimensions have
been proposed.  They provide an alternative view of the hierarchy
problem into geometric stabilization of the extra dimensions.

Collider experiments of the next generation will definitely search for signs
of various extra dimension models.  In this talk, I describe a few 
interesting physics that are quite unique for $\gamma\gamma$ colliders,
including 
large extra dimension model (ADD), warped extra dimension model 
(Randall-Sundrum), universal extra dimension model, as well as
techni-pions of technicolor models.

\section{ADD model}

It was proposed by Arkani, Dimopoulos, and Dvali \cite{arkani}
that the size $R$
of the extra dimensions that only gravity can propagate can be as
large as mm.  
Suppose the fundamental Planck scale of the model is $M_D$, the
observed Planck scale $M_{\rm Pl}$ becomes a derived quantity:
$
M^2_{\rm Pl} \sim M_D^{n+2} \; R^n \;.
$
If $R$ is extremely large, as large as a mm, the
fundamental Planck scale $M_D$ can be as low as TeV, which removes
the gauge hierarchy problem.
In this model, the SM particles and fields are confined to a brane while
only gravity is allowed to propagate in the extra dimensions.
Thus, the only probe of extra dimensions is through graviton interactions.
The graviton in the extra dimensions is equivalent to a tower of
Kaluza-Klein (KK) states in 4D point of view.
The separation between each state is of
order $1/R$, which is very small of order of $O(10^{-4})$ eV. This
means that in the energy scale of current high energy experiments,
the mass spectrum of the KK tower is effectively continuous.
Though each of the KK states interacts with a strength of
$1/M_{\rm Pl}$, however, when all the KK
states are summed up, the interaction has a strength of $1/M_D \sim 
1/O({\rm TeV})$.
 
Below $M_D$  the SM particles can
scatter into a graviton, which either (i) goes into extra
dimensions and does not come back to the brane, which then gives
rise to missing energy and momentum in experiments, or (ii) comes
back to the SM brane and decay back into SM particles, the
scattering amplitude of which then interferes with the normal SM
amplitude.  
Therefore, experimentally we can search for two types
of signatures, the missing energy or the interference effects.

\begin{figure}[t!]
\centering
\includegraphics[width=3in]{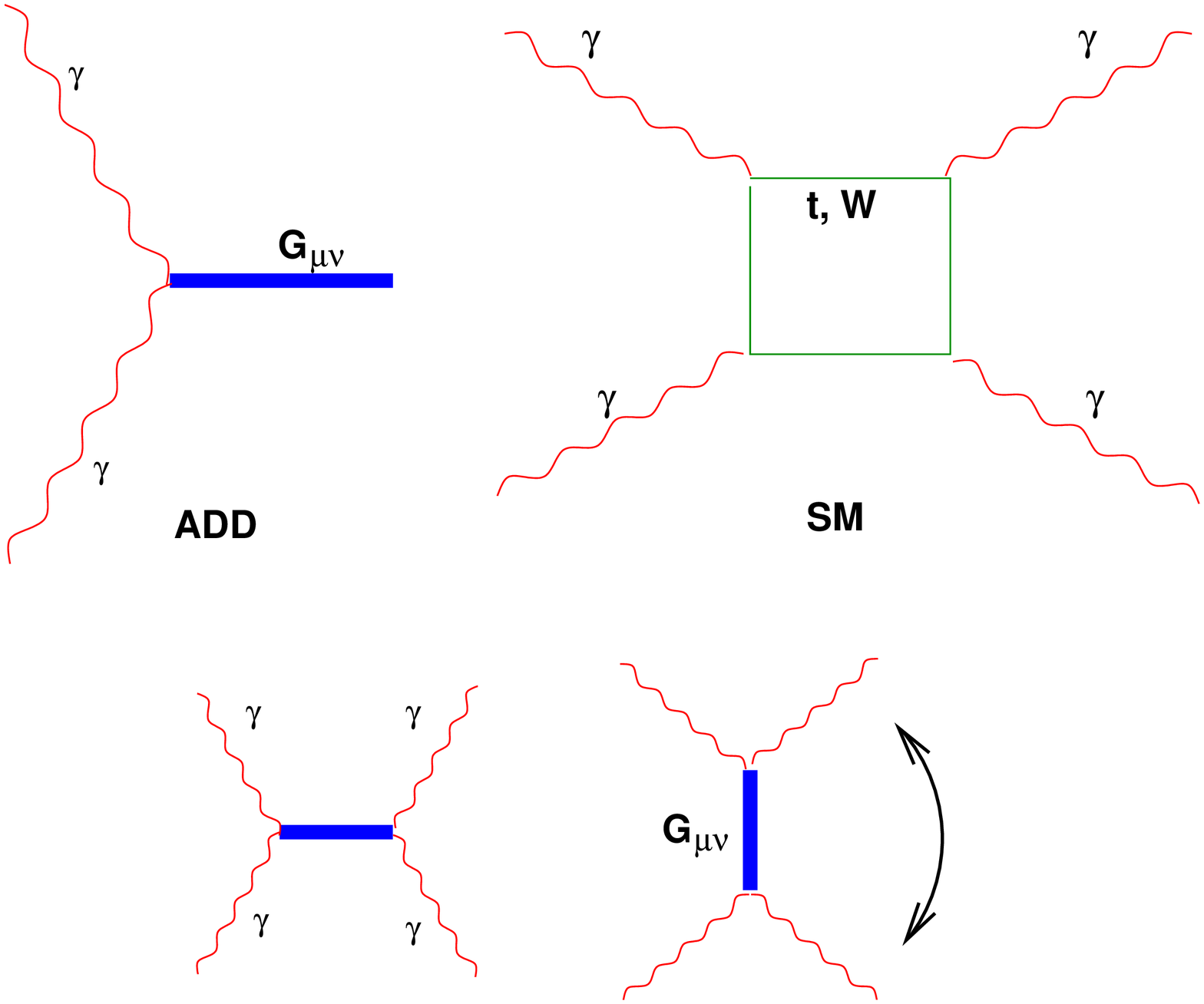}
\includegraphics[width=3in]{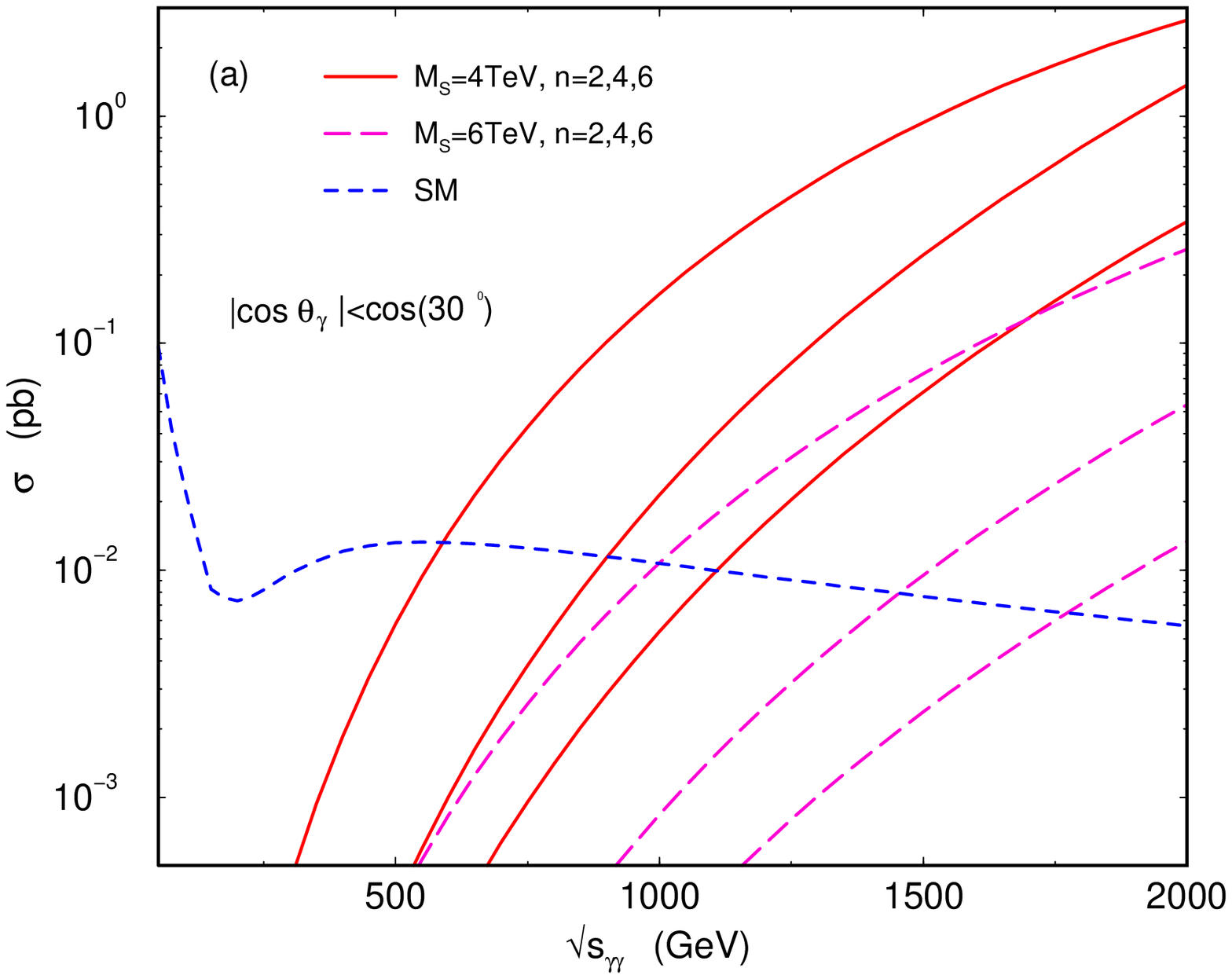}
\caption{\small \label{lbl}
(a) Feynman diagrams for $\gamma\gamma \to \gamma\gamma$ 
via graviton exchanges in $s$-, $t$-, and 
$u$-channels.  SM contributions occur through box diagrams.
(b) Cross sections for for $\gamma\gamma \to \gamma\gamma$ for graviton 
signals and SM background vs $\sqrt{s_{\gamma\gamma}}$
}
\end{figure}

\begin{figure}[t!]
\centering
\includegraphics[width=3in]{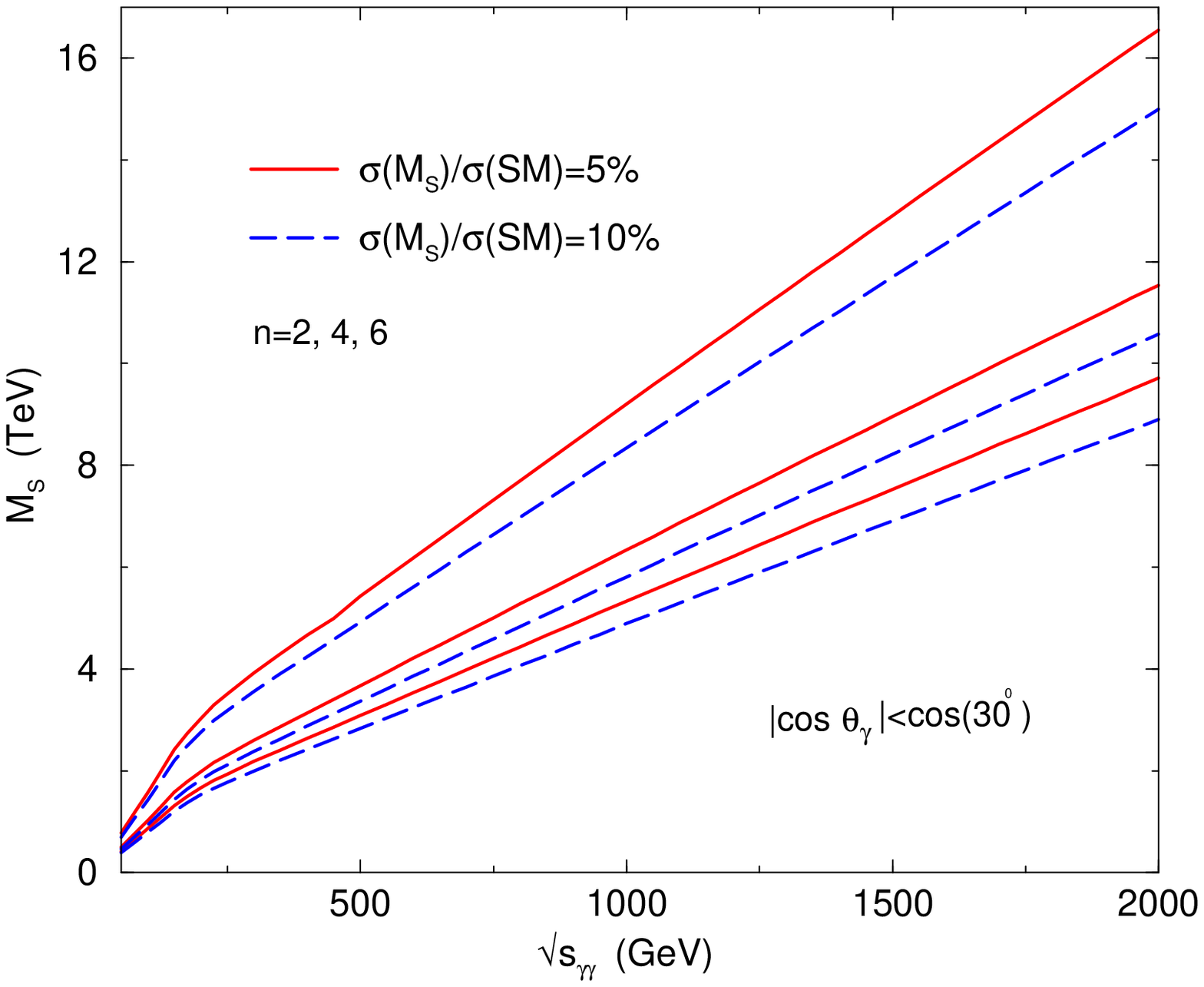}
\includegraphics[width=3in]{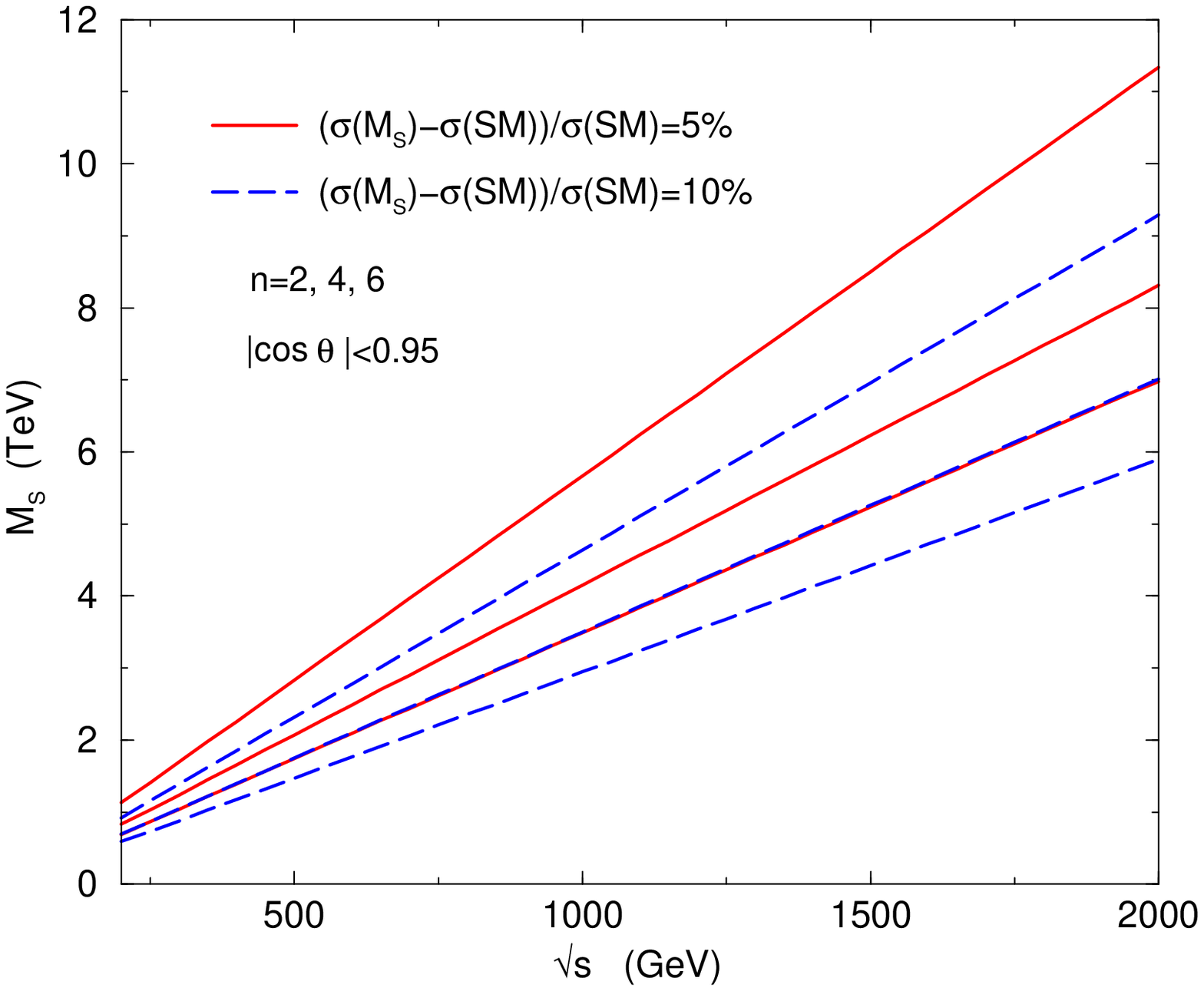}
\caption{\small \label{compare}
Comparison in sensitivity reach between $\gamma\gamma \to \gamma\gamma$ 
and $e^- e^+ \to \gamma\gamma$.}
\end{figure}

One unique process to search for virtual graviton effect is the 
light-by-light scattering \cite{add-11}.  The SM
amplitude has to go through a box diagram while the
graviton-induced amplitude is at tree-level: see Fig. \ref{lbl}(a).
In contrast, diphoton production at $pp$ or $e^+ e^-$ collisions
has tree-level SM contributions.  With graviton exchanges the differential
cross section is given by \cite{add-11}
\begin{equation}
\frac{d\sigma (\gamma\gamma \to \gamma\gamma)}{d |\cos\theta| } =
\frac{\pi s^3}{M_S^8}\, {\cal F}^2 \, \biggr [
1 + \frac{1}{8}\,(1+ 6 \cos^2\theta + \cos^4 \theta ) \biggr ] \;,
\end{equation}
where 
$ {\cal F} = \log (M_S^2/s)$ or $2/(n-2)$ for $n=2$ or $n>2$.
The cross sections for $\gamma\gamma \to \gamma\gamma$ for graviton 
signals and SM background vs $\sqrt{s_{\gamma\gamma}}$ are shown in 
Fig. \ref{lbl}(b).  It is clear that the signal can easily go above the SM
background when $\sqrt{s_{\gamma\gamma}} \agt 0.5$ TeV for $M_S\approx 4$ TeV.
We can also compare the sensitivity reach of $\gamma\gamma \to \gamma\gamma$ 
with $e^- e^+ \to \gamma\gamma$, as shown in Fig.~\ref{compare}.  
The latter has tree-level SM contributions.   
The $\gamma\gamma \to \gamma\gamma$ 
scattering can probe $M_S$ upto 16 TeV ($n=2$) 
for $\sqrt{s_{\gamma\gamma}}=2$ TeV
compared with about 11 TeV ($n=2$) by $e^- e^+ \to \gamma\gamma$.

\section{Randall-Sundrum model}

\begin{figure}[t!]
\centering
\includegraphics[width=3in]{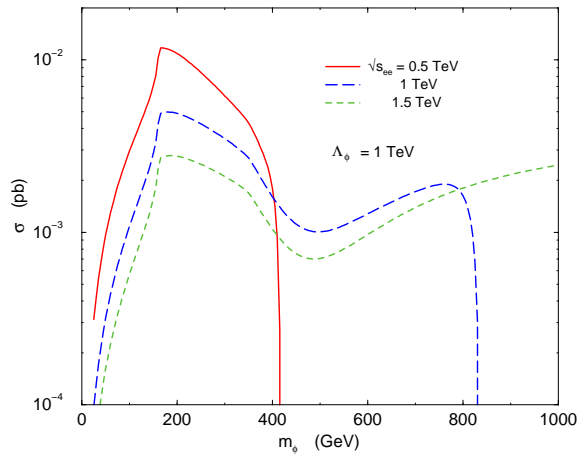}
\caption{\small \label{radion-cross}
Radion production vs radion mass.}
\end{figure}

The Randall-Sundrum (RS) model \cite{RS} beautifully explains the
gauge hierarchy with a moderate number through the exponential.
The most distinct feature of the RS model is the unevenly
spaced KK spectrum for the gravitons \cite{RS-1}.
Phenomenology
associated with the modulus field (known as the radion) \cite{RS-2},
describing the fluctuation in the separation of the two branes, is
another interesting feature of the RS model.
The interactions of the radion with SM particles are given by
\begin{equation}
{\cal L}_{\rm int} = \frac{\phi}{\Lambda_\phi} \; T^\mu_\mu ({\rm
SM}) \;,
\end{equation}
where $\Lambda_\phi= \langle \phi \rangle$ is of order of TeV and
$T^\mu_\mu ({\rm SM})$ is the trace of the energy-momentum tensor.
It is clear that the interactions are very similar to those of the
SM Higgs boson with the replacement of the VEV by $\langle \phi \rangle$.
However, the radion has anomalous couplings to gluons and photons
from the trace anomaly:
\[
T^\mu_\mu({\rm SM})^{\rm anom} = \sum_a \frac{\beta_a (g_a)}{2g_a}
F_{\mu\nu}^a F^{a \mu\nu} \;,
\]
where $\beta_a$ are the beta functions for the gauge groups.
Because of the anomalous coupling of the radion to photons, 
photon fusion would be an important production channel for the
radion in $\gamma\gamma$ collisions.
In Fig.~\ref{radion-cross}, we show the production cross section of the
radion at photon colliders \cite{RS-6}.

Another interesting feature of the radion is the possibility of mixing
between the radion and the Higgs boson \cite{RS-4}.
This is the consequence of a mixing term in the action
\[
S_\xi = \xi \int d^4 x \sqrt{g_{\rm vis} } R(g_{\rm vis}) \hat H^\dagger 
\hat H \;,
\]
where $R(g_{\rm vis})$ is the Ricci scalar on the visible brane.  A nonzero
$\xi$ will induce some special triple vertices \cite{RS-7}:
$
h\,\mbox{-}\,\phi\,\mbox{-}\,\phi, \;
h_{\mu\nu}^{(n)}\,\mbox{-}\,h\,\mbox{-}\,\phi, \;
\phi\,\mbox{-}\,\phi\,\mbox{-}\,\phi,\;
h_{\mu\nu}^{(n)}\,\mbox{-}\,\phi\,\mbox{-}\,\phi  \;.
$
One can then use 
\[
\gamma\gamma \to G_{\mu\nu}^{(n)} \to h\phi
\]
to probe the radion-Higgs mixing.  Similar studies \cite{song} have
been performed for $pp$ and $e^+ e^-$ colliders.

\section{Universal Extra Dimensions}

In this scenario, all SM particles are free to move in all dimensions, 
dubbed universal extra dimensions \cite{ue-1}.  
Consider the case with only one extra dimension.
The momentum conservation in the fifth dimension, after compactification, 
becomes conservation in KK numbers.
There may be some boundary terms arising
from the fixed points that break the conservation of KK numbers into a $Z_2$ 
parity, called KK parity.  Odd parities are assigned to 
the KK states with an odd KK number.  SM particles have even KK
parity.  The lightest KK state is often the first KK state of the
hypercharge gauge boson $\gamma^{(1)}$.  The consequence of the KK parity is
that this $\gamma^{(1)}$ becomes stable and gives missing energy signal.
Collider phenomenology is mainly the pair production of KK quarks and 
KK leptons.
\[
\gamma\gamma \to q^{(1)} q'^{(1)}, \; \ell^{+(1)} \ell^{-(1)} \;.
\]
Each KK quark decays into a light quark and $\gamma^{(1)}$ while each
KK lepton decays into a lepton and $\gamma^{(1)}$.  Therefore, in the final
state there are multi-jets or multi-leptons plus missing energy.

\section{Techni-Pions}
\begin{figure}[t!]
\centering
\includegraphics[width=3in]{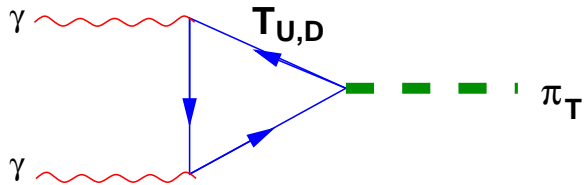}
\caption{\small \label{anomaly}
Anomaly coupling of the $\pi^0_T$ to a pair of photons}
\end{figure}

Because of the anomaly-type coupling, such as in QCD $\pi^0\gamma\gamma$,
we can use $\gamma\gamma$ collisions to probe for any QCD-like $\pi^0$
resonances.  Such resonances are often predicted in some strongly-interacting
models, e.g., technicolor models \cite{tc}.  
We use the Technicolor Straw Man model \cite{tc} as illustration.  
The lightest techni-mesons are constructed solely from the lightest
techni-fermion doublet $(T_U, T_D)$, from which isotriplet 
$\rho_T^{0,\pm}, \pi_T^{0,\pm}$ and isosinglet $\pi^{'0}_T, \, \omega^0_T$
can be formed.  In particular, the neutral $\pi^0_T$ and $\pi^{'0}_T$
have an anomaly coupling, shown in Fig.~\ref{anomaly}:
\[
\Gamma = N_{TC} {\cal V}_{\gamma\gamma \pi} \frac{e^2}{8\pi^2 F_T}\,
 \epsilon^{\mu\nu\lambda \rho} \epsilon^*_{1\mu} \epsilon^*_{2\nu} p_{1\lambda}
p_{2\rho} \;.
\]
Preliminary calculation shows that the cross section is \cite{tseng}
\[
\sigma = \int \frac{d x}{x} \, f_{\gamma/e}(x) f_{\gamma/e}(m_{\pi_T}^2/s x)\,
N_{TC}^2 V_{\gamma\gamma\pi}^2 \frac{\alpha^2}{32 \pi F_T^2} \,
 \frac{m_{\pi_T}^2}{s} \; \sim O(1 {\rm fb}) \;.
\]
With the SM background coming only from box diagrams, this signal has a good
chance to be observed.

In summary, I have described some interesting new physics that are quite
unique to photon colliders.  There are of course many other interesting
physics that can be performed.  For some reviews
please refer to Refs. \cite{review}.

\section*{Acknowledgments}
This research was supported in part by
the National Science Council of Taiwan R.O.C. under grant no.
NSC 92-2112-M-007-053- and 93-2112-M-007-025-.

\end{document}